\documentclass[twocolumn,showpacs,preprintnumbers,amsmath,amssymb,prl]{revtex4}

\usepackage{graphicx}
\usepackage{dcolumn}
\usepackage{bm}
\usepackage{tabularx}
\usepackage{amsmath}

\begin{document}

\title{Nodeless superconducting gap in electron-doped BaFe$_{1.9}$Ni$_{0.1}$As$_2$ probed by quasiparticle heat transport}

\author{L. Ding,$^1$ J. K. Dong,$^1$ S. Y. Zhou,$^1$ T. Y. Guan,$^1$ X. Qiu,$^1$ \\ C. Zhang,$^1$ L. J. Li,$^2$ X. Lin,$^2$ G. H. Cao,$^2$ Z. A. Xu,$^2$ S. Y. Li$^{1,*}$}

\affiliation{$^1$Department of Physics, Surface Physics Laboratory (National Key Laboratory), and Laboratory of Advanced Materials, Fudan University, Shanghai 200433, P. R. China\\
$^2$Department of Physics, Zhejiang University, Hangzhou 310027, P.
R. China}

\date{\today}

\begin{abstract}
The in-plane thermal conductivity $\kappa$ of electron-doped
iron-arsenide superconductor BaFe$_{1.9}$Ni$_{0.1}$As$_2$ ($T_c$ =
20.3 K) single crystal was measured down to 70 mK. In zero field,
the absence of a residual linear term $\kappa_0/T$ at $ T
\rightarrow 0$ is strong evidence for nodeless superconducting gap.
In magnetic field, $\kappa_0/T$ shows a slow field dependence up to
$H$ = 14.5 T ($\approx$ 30\% $H_{c_2}$). This is consistent with the
superconducting gap structure demonstrated by angle-resolved
photoemission spectroscopy experiments in
BaFe$_{1.85}$Co$_{0.15}$As$_2$ ($T_c$ = 25.5 K), where isotropic
superconducting gaps with similar size on hole and electron pockets
were observed.
\end{abstract}


\maketitle

The recent discovery of iron-based superconductors with $T_c$ as
high as 55 K \cite{Kamihara,XHChen,GFChen,ZARen,RHLiu} has attracted
great attention. As a second family of high temperature
superconductors after cuprates, the pairing symmetry of its
superconducting gap is one of the most important issue to address.
Spin triplet pairing was first ruled out in
BaFe$_{1.8}$Co$_{0.2}$As$_2$ ($T_c$ = 22 K) single crystal by the
NMR Knight shift measurements \cite{FLNing}. This leaves three
possible singlet paring candidates: conventional $s$, $d$, and
$s_{\pm}$, a superconducting state with order parameters of opposite
signs on the electron and hole pockets \cite{Mazin}. While Andreev
spectroscopy \cite{TYChen}, angle-resolved photoemission
spectroscopy (ARPES)
\cite{HDing,TKondo,LZhao,LWray,KNakayama,KTerashima}, and latest
specific heat \cite{GMu} experiments on FeAs-superconductors support
full superconducting gaps without nodes, NMR data
\cite{KMatano,HGrafe,MYashima} and extensive penetration depth
studies \cite{CMartin1,KHashimoto,LMalone,CMartin2,RGordon} reveal a
contradictory picture of either nodeless or nodal superconductivity.
Even if the nodeless superconducting gap is eventually confirmed,
clear-cut experiments to distinguish $s_{\pm}$ from conventional
$s$-wave have to be done. Therefore, the paring symmetry in
iron-arsenide superconductors is still far from consensus.

Low-temperature thermal conductivity measurement is a powerful bulk
tool to probe the superconducting gap structure \cite{Shakeripour}.
For unconventional superconductors with nodes in the superconducting
gap, like $d$-wave cuprates and $p$-wave ruthenate, the nodal
quasiparticles will contribute a finite $\kappa_0/T$ in zero field
\cite{Proust,Suzuki}. So far, only one heat transport study was
reported for FeAs-based superconductors \cite{XGLuo}. For the
hole-doped Ba$_{1-x}$K$_x$Fe$_2$As$_2$ ($T_c \simeq$ 30 K) single
crystal, a negligible $\kappa_0/T$ was found in zero field,
indicating a full superconducting gap. However, $\kappa_0/T$
increases rapidly with magnetic field even for $H \ll H_{c_2}$,
which was inferred that the gap must be very small on some portion
of the Fermi surface, whether from strong anisotropy or band
dependence, or both. To clarify this important issue, more heat
transport experiments on other FeAs-based superconductors are
needed.

In this Letter, we probe the superconducting gap of electron-doped
BaFe$_{1.9}$Ni$_{0.1}$As$_2$ by measuring the thermal conductivity
$\kappa$ of a single crystal with $T_c = 20.3$ K down to 70 mK. In
zero field, the residual linear term $\kappa_0/T$ is negligible, a
clear indication that BaFe$_{1.9}$Ni$_{0.1}$As$_2$ has nodeless
superconducting gap. In magnetic field, $\kappa_0/T(H)$ shows a slow
field dependence, different from the case of hole-doped
Ba$_{1-x}$K$_x$Fe$_2$As$_2$. This difference is discussed on the
base of superconducting gap structure in these two systems measured
by ARPES.

Single crystals with nominal formula BaFe$_{1.9}$Ni$_{0.1}$As$_2$
were prepared by self flux method \cite{LJLi}. Energy Dispersive of
X-ray (EDX) microanalysis show that the actual Ni content is 0.096,
close to the nominal composition. The ac magnetization was measured
in a Quantum Design Physical Property Measurement System (PPMS). The
sample was cleaved to a rectangular shape of dimensions 1.5 $\times$
0.88 mm$^2$ in the plane, with 55 $\mu$m thickness along the
$c$-axis. Contacts were made directly on the fresh sample surfaces
with silver paint, which were used for both resistivity and thermal
conductivity measurements. The contacts are metallic with typical
resistance 50 m$\Omega$ at 1.5 K. In-plane thermal conductivity was
measured in a dilution refrigerator down to 70 mK, using a standard
four-wire steady-state method with two RuO$_2$ chip thermometers,
calibrated {\it in situ} against a reference RuO$_2$ thermometer.
Magnetic fields were applied along the $c$-axis and perpendicular to
the heat current. To ensure a homogeneous field distribution in the
sample, all fields were applied at temperature above $T_c$.

Fig. 1a shows the in-plane resistivity of our
BaFe$_{1.9}$Ni$_{0.1}$As$_2$ single crystal in zero field. The
middle point of the resistive transition is at $T_c$ = 20.3 K, in
good agreement with previous study \cite{LJLi}. The 10-90\% width of
the resistive transition is less than 0.3 K, indicating the high
homogeneity of our crystal. The residual resistivity $\rho_0$ = 132
$\mu \Omega$ cm is extrapolated from the data above $T_c$ by using
the Fermi liquid form $\rho = \rho_0 + AT^2$. In Fig. 1b, the
normalized ac magnetization also shows a sharp superconducting
transition similar to Fig. 1a.

\begin{figure}
\includegraphics[clip,width=6cm]{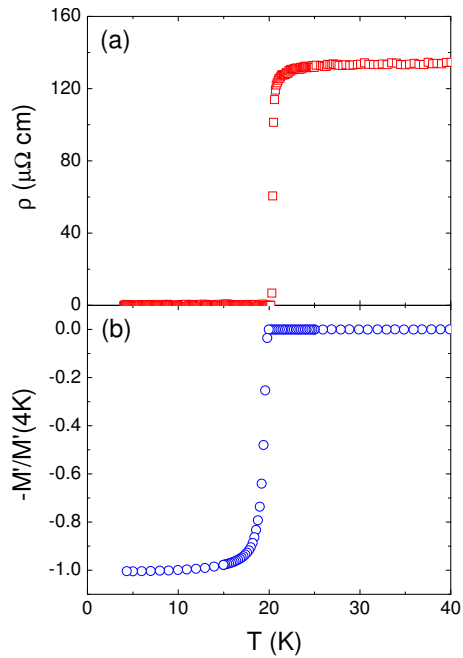}
\caption{(Color online) (a) In-plane resistivity $\rho(T)$ of
BaFe$_{1.9}$Ni$_{0.1}$As$_2$ single crystal. The middle point of the
resistive transition is at $T_c$ = 20.3 K, and the 10-90\% width of
the transition is less than 0.3 K. (b) Normalized ac magnetization.}
\end{figure}

In Fig. 2, the temperature dependence of the in-plane thermal
conductivity for BaFe$_{1.9}$Ni$_{0.1}$As$_2$ in zero field is
plotted as $\kappa / T$ vs $T$. Since both electrons and phonons
contribute to the measured conductivity, we fit the data to
$\kappa/T = a + bT^{\alpha-1}$ \cite{Sutherland,SYLi}, where $aT$
and $bT^{\alpha}$ represent electronic and phonon contributions,
respectively. For phonon scattering off the crystal bourdary at low
temperature, one usually gets $\alpha = 3$, but specular reflection
of phonons at the smooth crystal surfaces can result in a lower
power $\alpha < 3$ \cite{Sutherland,SYLi}. For
BaFe$_{1.9}$Ni$_{0.1}$As$_2$, it is found that the data below 0.8 K
can be well fitted (the solid line in Fig. 2) and gives $\kappa_0/T$
= -3 $\pm$ 2 $\mu$W K$^{-2}$ cm$^{-1}$, with $\alpha$ = 2.02 $\pm$
0.01.

In the non-superconducting parent BaFe$_2$As$_2$ single crystal, the
Wiedemann-Franz law, which relates charge and thermal conductivities
by $\kappa/T = L_0/\rho$ with $L_0$ the Lorenz number 2.45 $\times
10^{-8}$ W $\Omega$ K$^{-2}$, was found to be satisfied as $T
\rightarrow 0$ \cite{Kurita2}. For BaFe$_{1.9}$Ni$_{0.1}$As$_2$ with
$T_c$ = 20.3 K, the normal-state Wiedemann-Franz law expectation is
$\kappa_0/T = L_0/\rho_0$ = 0.186 mW K$^{-2}$ cm$^{-1}$ with
$\rho_0$ = 132 $\mu \Omega$ cm, the dashed line in Fig. 2.

Since the residual linear term $\kappa_0/T$ is within the
experimental error bar $\pm$ 5 $\mu$W K$^{-2}$ cm$^{-1}$
\cite{SYLi}, which is less than 3\% of the normal-state value, the
electronic contribution to the thermal conductivity is negligible in
zero field. This is consistent with previous results on hole-doped
Ba$_{1-x}$K$_x$Fe$_2$As$_2$ single crystals \cite{XGLuo} and the
low-$T_c$ superconductor BaNi$_2$As$_2$ ($T_c$ = 0.7 K)
\cite{Kurita1}, suggesting a nodeless (at least in $ab$-plane)
superconducting gap. However, the power $\alpha = 2.02$ of the
phonon conductivity $bT^{\alpha}$ is much lower than $\alpha = 2.65$
found in Ba$_{1-x}$K$_x$Fe$_2$As$_2$ \cite{XGLuo}. We note that in
the parent compound BaFe$_2$As$_2$ single crystal \cite{Kurita2},
the power $\alpha = 2.22$ is more closer to our value. Whether
specular reflections of the phonon boundary scattering
\cite{Sutherland,SYLi} can give such a low $\alpha$ is not clear to
us. In fact, phonons scattering off either electrons or grain
boundaries will give $\alpha = 2$ \cite{Berman}. Therefore, more
experimental results are needed to clarify the temperature
dependence of phonon thermal conductivity in FeAs-compound single
crystals.

\begin{figure}
\includegraphics[clip,width=7.5cm]{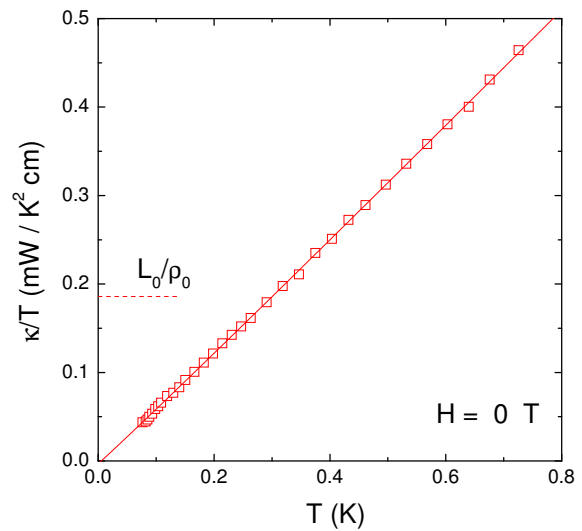}
\caption{(Color online) Temperature dependence of the in-plane
thermal conductivity for BaFe$_{1.9}$Ni$_{0.1}$As$_2$ single crystal
in zero field. The solid line represents a fit of the data to
$\kappa/T = a + bT^{\alpha-1}$. This gives the residual linear term
$\kappa_0/T$ = -3 $\pm$ 2 $\mu$W K$^{-2}$ cm$^{-1}$, which is
negligible within the experimental error bar. The dashed line is the
normal-state Wiedemann-Franz law expectation at $T \rightarrow 0$,
namely $L_0$/$\rho_0$, with $L_0$ the Lorenz number 2.45 $\times
10^{-8}$ W $\Omega$ K$^{-2}$.}
\end{figure}

Fig. 3 shows the low-temperature thermal conductivity of
BaFe$_{1.9}$Ni$_{0.1}$As$_2$ in magnetic fields applied along the
$c$-axis ($H$ = 0, 9, 13, and 14.5 T). The data of $\kappa/T$ in
high fields below 0.25 K manifests similar temperature dependence to
the zero field data. We fit the $H$ = 9, 13, and 14.5 T curves by
using the same equation $\kappa/T = a + bT^{\alpha-1}$, with fixed
$\alpha = 2.02$ due to the slighly increasing noise level of the
in-field data. The solid lines are the fitting curves, which give
$\kappa_0/T$ = 4, 15, and 20 $\mu$W K$^{-2}$ cm$^{-1}$ for $H$ = 9,
13, and 14.5 T, respectively.

The upper critical field $H_{c2}$ of BaFe$_{1.9}$Ni$_{0.1}$As$_2$
($T_c$ = 20.3 K) single crystal has not been determined yet. For
BaFe$_{1.8}$Co$_{0.2}$As$_2$ ($T_c$ = 22 K) single crystal, the
$H_{c2}$ was estimated $\sim$ 50 T \cite{AYamamoto}. Taking this
value as the $H_{c2}$ of our BaFe$_{1.9}$Ni$_{0.1}$As$_2$ sample,
$H$ = 14.5 T is just about 30\% of $H_{c2}$.

In Fig. 4, the normalized $\kappa_0/T$ of
BaFe$_{1.9}$Ni$_{0.1}$As$_2$ is plotted as a function of $H/H_{c2}$,
together with the clean $s$-wave superconductor Nb \cite{Lowell},
the dirty $s$-wave superconducting alloy InBi \cite{Willis}, the
multi-band $s$-wave superconductor NbSe$_2$ \cite{Boaknin}, an
overdoped sample of the $d$-wave superconductor Tl-2201
\cite{Proust}, and Ba$_{0.75}$K$_{0.25}$Fe$_2$As$_2$ \cite{XGLuo}.
For a clean (like Nb) or dirty (like InBi) type-II $s$-wave
superconductor with isotropic gap, $\kappa_0/T$ should grow
exponentially with field (above $H_{c1}$). This usually gives
negligible $\kappa_0/T$ for field lower than $H_{c2}/4$, as seen in
Fig. 4. For NbSe$_2$, $\kappa_0/T$ increases much rapid at low
field. This can be explained by its multi-gap structure, whereby the
gap on the $\Gamma$ band is approximately one third of the gap on
the other two Fermi surfaces, and magnetic field will first
suppresses the superconductivity on the Fermi surface with smaller
gap (given that $H_{c2}(0) \propto \Delta_0^2$) \cite{Boaknin}.

As seen in Fig. 4, the $\kappa_0/T(H)$ of
BaFe$_{1.9}$Ni$_{0.1}$As$_2$ more likely follows the behavior of
isotropic $s$-wave gap. This field dependence is different from that
of the hole-doped Ba$_{1-x}$K$_x$Fe$_2$As$_2$ ($T_c \simeq$ 30 K)
sample \cite{XGLuo}, where $\kappa_0/T$ increases almost linearly
with $H$ up to 15 T. Such a rapid increase of $\kappa_0/T(H)$ in
Ba$_{1-x}$K$_x$Fe$_2$As$_2$ has been interpreted as evidence for a
$k$-dependent gap magnitude, coming from angle ({\it i.e.},
anisotripic) or band ({\it i.e.}, isotropic but with different
magnetitudes on different bands) dependence, or both \cite{XGLuo}.

\begin{figure}
\includegraphics[clip,width=7.5cm]{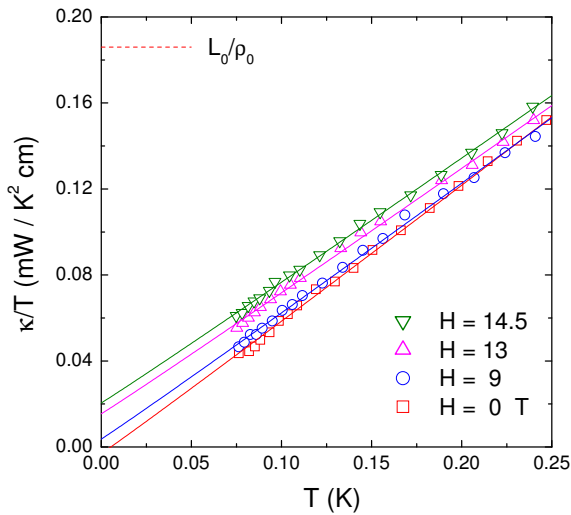}
\caption{(Color online) Low-temperature thermal conductivity of
BaFe$_{1.9}$Ni$_{0.1}$As$_2$ in magnetic fields applied along the
$c$-axis ($H$ = 0, 4, 9, and 14.5 T). The solid lines are $\kappa/T
= a + bT^{\alpha-1}$ fits (see text). The dashed line is the normal
state Wiedemann-Franz law expectation $L_0$/$\rho_0$.}
\end{figure}

\begin{figure}
\includegraphics[clip,width=7.2cm]{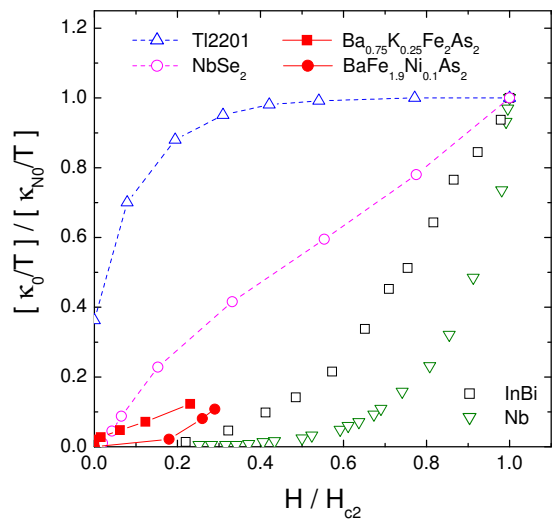}
\caption{(Color online) Normalized residual linear term $\kappa_0/T$
of BaFe$_{1.9}$Ni$_{0.1}$As$_2$ as a function of $H/H_{c2}$. Similar
data of the clean $s$-wave superconductor Nb \cite{Lowell}, the
dirty $s$-wave superconducting alloy InBi \cite{Willis}, the
multi-band $s$-wave superconductor NbSe$_2$ \cite{Boaknin}, an
overdoped sample of the $d$-wave superconductor Tl-2201
\cite{Proust}, and Ba$_{0.75}$K$_{0.25}$Fe$_2$As$_2$ \cite{XGLuo}
are also shown for comparison.}
\end{figure}

In order to explain this difference, let us examine the gap values
on all Fermi surface (FS) sheets for both hole- and electron-doped
BaFe$_2$As$_2$ measured by ARPES \cite{HDing,KNakayama,KTerashima}.
For hole-doped Ba$_{0.6}$K$_{0.4}$Fe$_2$As$_2$ ($T_c$ = 37 K), the
average gap values $\Delta(0)$ for the two hole ($\alpha$ and
$\beta$) pockets are 12.5 and 5.5 meV, respectively, while for the
electron ($\gamma$ and $\delta$) pockets, the gap value is about
12.5 meV \cite{HDing,KNakayama}. For electron-doped
BaFe$_{1.85}$Co$_{0.15}$As$_2$ ($T_c$ = 25.5 K), the average gap
values $\Delta(0)$ of hole ($\beta$) and electron ($\gamma$ and
$\delta$) pockets are 6.6 and 5.0 meV, respectively
\cite{KTerashima}.

Since the doping level and $T_c$ of our BaFe$_{1.9}$Ni$_{0.1}$As$_2$
sample are close to those of BaFe$_{1.85}$Co$_{0.15}$As$_2$, their
superconducting gap structure should also be similar. Therefore, due
to the similar sizes (6.6 vs 5.0 meV) of these isotropic
superconducting gaps, the $\kappa_0/T(H)$ of our
BaFe$_{1.9}$Ni$_{0.1}$As$_2$ sample behaves more like a conventional
single-gap $s$-wave superconductor. For hole-doped
Ba$_{0.6}$K$_{0.4}$Fe$_2$As$_2$, the sizes of these gaps are quiet
different (12.5 vs 5.5 meV), which gives a ratio $R$ = 12.5/5.5 =
2.3 \cite{HDing,KNakayama}. Taking this ratio for the slightly
underdoped Ba$_{1-x}$K$_x$Fe$_2$As$_2$ \cite{XGLuo}, it is smaller
than that in NbSe$_2$ ($R$ $\approx$ 3). This may explains the
nearly linear increase of $\kappa_0/T(H)$ in
Ba$_{1-x}$K$_x$Fe$_2$As$_2$ with the slope smaller than that in
NbSe$_2$ \cite{XGLuo}, given $H_{c2}(0) \propto \Delta_0^2$ and
magnetic field first suppresses the superconductivity on the Fermi
surface with smallest superconducting gap.

In summary, we have used low-temperature thermal conductivity to
clearly demonstrate nodeless superconducting gap in electron-doped
iron-arsenide superconductor BaFe$_{1.9}$Ni$_{0.1}$As$_2$.
Furthermore, the $\kappa_0/T(H)$ shows a slow $H$ dependence at low
field, different from the rapid, linear $\kappa_0/T(H)$ in
hole-doped Ba$_{1-x}$K$_x$Fe$_2$As$_2$. This difference can be
explained by the different ratio of the band-dependent
superconducting gaps. Our results are consistent with nodeless
multi-gaps in iron-arsenide superconductors, as revealed by ARPES
experiments.

{\it Note:} After our present work first appeared on arXiv
(0906.0138), two similar works on BaFe$_{2-x}$Co$_x$As$_2$ were also
put on arXiv \cite{YMachida,Tanatar}. In Ref. [38], a large
$\kappa_0/T$ was observed in BaFe$_{1.86}$Co$_{0.14}$As$_2$ single
crystal at zero field, apparently contradicting to our results. In
Ref. [39], the results of BaFe$_{2-x}$Co$_x$As$_2$ with $x$ = 0.148
are consistent with ours.

This work is supported by the Natural Science Foundation of China,
the Ministry of Science and Technology of China (National Basic
Research Program No:2009CB929203 and 2006CB601003), Program for New
Century Excellent Talents in University,
and STCSM of China (No: 08dj1400200 and 08PJ1402100).\\

$^*$ E-mail: shiyan$\_$li@fudan.edu.cn


\begin{thebibliography}{99}

\bibitem{Kamihara} Y. Kamihara {\it et al.}, J. Am. Chem. Soc. {\bf 130}, 3296 (2008).
\bibitem{XHChen} X. H. Chen {\it et al.}, Nature {\bf 453}, 761 (2008).
\bibitem{GFChen} G. F. Chen {\it et al.}, Phys. Rev. Lett. {\bf 100}, 247002 (2008).
\bibitem{ZARen} Z. A. Ren {\it et al.}, Chin. Phys. Lett. {\bf 25}, 2215 (2008).
\bibitem{RHLiu} R. H. Liu {\it et al.}, Phys. Rev. Lett. {\bf 101}, 087001 (2008).
\bibitem{FLNing} F. L. Ning {\it et al.}, J. Phys. Soc. Jap. {\bf 77}, 103705 (2008).
\bibitem{Mazin} I. I. Mazin {\it et al.}, Phys. Rev. Lett. {\bf 101}, 057003 (2008).
\bibitem{TYChen} T. Y. Chen {\it et al.}, Nature {\bf 453}, 1224 (2008).
\bibitem{HDing} H. Ding {\it et al.}, Europhys. Lett. {\bf 83} 47001 (2008).
\bibitem{TKondo} T. Kondo {\it et al.}, Phys. Rev. Lett. {\bf 101}, 147003 (2008).
\bibitem{LZhao} L. Zhao {\it et al.}, Chin. Phys. Lett. {\bf 25} 4402 (2008).
\bibitem{LWray} L. Wray {\it et al.}, Phys. Rev. B {\bf 78} 184508 (2008).
\bibitem{KNakayama} K. Nakayam {\it et al.}, arXiv:0812.0663.
\bibitem{KTerashima} K. Terashima {\it et al.}, arXiv:0812.3704.
\bibitem{GMu} G. Mu {\it et al.}, Phys. Rev. B {\bf 79}, 174501 (2009).
\bibitem{KMatano} K. Matano {\it et al.}, Europhys. Lett. {\bf 83} 57001 (2008).
\bibitem{HGrafe} H. -J. Grafe {\it et al.}, Phys. Rev. Lett. {\bf 101}, 047003 (2008).
\bibitem{MYashima} M. Yashima {\it et al.}, arXiv:0905.1896.
\bibitem{CMartin1} C. Martin {\it et al.}, arXiv:0807.0876.
\bibitem{KHashimoto} K. Hashimoto {\it et al.}, Phys. Rev. Lett. {\bf 102}, 017002 (2009).
\bibitem{LMalone} L. Malone {\it et al.}, arXiv:0806.3908.
\bibitem{CMartin2} C. Martin {\it et al.}, arXiv:0901.1804.
\bibitem{RGordon} R. T. Gordon {\it et al.}, Phys. Rev. Lett. {\bf 102}, 127004 (2009).
\bibitem{Shakeripour} H. Shakeripour {\it et al.}, New Journal
of Physics {\bf 11}, 055065 (2009).
\bibitem{Proust} C. Proust {\it et al.}, Phys. Rev. Lett. {\bf 89}, 147003 (2002).
\bibitem{Suzuki} M. Suzuki {\it et al.}, Phys. Rev. Lett. {\bf 88}, 227004 (2002).
\bibitem{XGLuo} X. G. Luo {\it et al.}, arXiv:0904.4049.
\bibitem{LJLi} L. J. Li {\it et al.}, New J. Phys. {\bf 11}, 025008 (2009).
\bibitem{Sutherland} M. Sutherland {\it et al.}, Phys. Rev. B {\bf 67}, 174520 (2003).
\bibitem{SYLi} S. Y. Li {\it et al.}, Phys. Rev. B {\bf 77}, 134501 (2008).
\bibitem{Kurita2} N. Kurita {\it et al.}, arXiv:0904.4470.
\bibitem{Kurita1} N. Kurita {\it et al.}, Phys. Rev. Lett. {\bf 102}, 147004 (2009).
\bibitem{Berman} R. Berman, {\it Thermal conduction in
Solids} (Oxford Univ. Press, Oxford) (1976).
\bibitem{AYamamoto} A. Yamamoto {\it et al.}, Appl. Phys. Lett. {\bf 94}, 062511 (2009).
\bibitem{Lowell} J. Lowell and J. Sousa, J. Low. Temp. Phys. {\bf 3}, 65 (1970).
\bibitem{Willis} J. Willis and D. Ginsberg, Phys. Rev. B {\bf 14}, 1916 (1976).
\bibitem{Boaknin} E. Boaknin {\it et al.}, Phys. Rev. Lett. {\bf 90}, 117003 (2003).
\bibitem{YMachida} Y. Machida {\it et al.}, arXiv:0906.0508.
\bibitem{Tanatar} M. A. Tanatar {\it et al.}, arXiv:0907.1276.

\end{thebibliography}
\end{document}